\documentclass[reprint, superscriptaddress,
nofootinbib,
bibnotes,
amsmath,amssymb,
aps,
prb,
]{revtex4-2}

\usepackage{graphicx}
\usepackage{dcolumn}
\usepackage{bm}
\usepackage{hyperref}
\usepackage[mathlines]{lineno}
\usepackage{xcolor}
\usepackage{threeparttable}
\usepackage{booktabs}
\usepackage{float} 

\begin{document}
\renewcommand{\thefigure}{\arabic{figure}}
\preprint{APS/123-QED}

\title{From Bloch to Néel: Anisotropy-dependent Domain-Wall Character in FePd Thin Films}
\author{Annika Stellhorn}\email[Corresponding author: ]{annika.stellhorn@ess.eu}
\affiliation{European Spallation Source ERIC, Partikelgatan 2, 224 84 Lund, Sweden}
\affiliation{Division of Synchrotron Radiation Research, Lund University, 22100 Lund, Sweden}
\author{Alicia Backs}
\affiliation{European Spallation Source ERIC, Partikelgatan 2, 224 84 Lund, Sweden} 
\author{Ekaterina Klyushina}
\affiliation{Division of Synchrotron Radiation Research, Lund University, 22100 Lund, Sweden}
\author{Connie Bednarski-Meinke}
\affiliation{J{\"u}lich Centre for Neutron Science (JCNS-2), Forschungszentrum Jülich GmbH, 52425 Jülich, Germany}
\author{Steffen Tober}
\affiliation{J{\"u}lich Centre for Neutron Science (JCNS-2), Forschungszentrum Jülich GmbH, 52425 Jülich, Germany}
\author{Denis M. Vasiukov}
\affiliation{European Spallation Source ERIC, Partikelgatan 2, 224 84 Lund, Sweden}
\affiliation{Division of Synchrotron Radiation Research, Lund University, 22100 Lund, Sweden}
\author{Oskar Stepančić}
\affiliation{Division of Synchrotron Radiation Research, Lund University, 22100 Lund, Sweden}
\author{Vítor Alexandre de Oliveira Lima}
\affiliation{J{\"u}lich Centre for Neutron Science (JCNS-2), Forschungszentrum Jülich GmbH, 52425 Jülich, Germany} 
\author{Lukas Aniansson}
\affiliation{Division of Synchrotron Radiation Research, Lund University, 22100 Lund, Sweden}
\author{Paul Steadman}
\affiliation{Diamond Light Source, Didcot, Oxfordshire, United Kingdom}
\author{Manuel Valvidares}
\affiliation{ALBA Synchrotron, Cerdanyola del Vallès, Barcelona, Spain}
\author{J\"org Schwenke}
\affiliation{MAX IV Laboratory, Lund University, P O Box 118, 22100 Lund, Sweden}
\author{Claudiu Bulbucan}
\affiliation{MAX IV Laboratory, Lund University, P O Box 118, 22100 Lund, Sweden}
\author{Elizabeth Blackburn}
\affiliation{Division of Synchrotron Radiation Research, Lund University, 22100 Lund, Sweden}

\begin{abstract}

We report an experimental investigation of the depth-dependent chiral magnetic domain wall texture in L1$_0$-FePd thin films with high perpendicular magnetic anisotropy. Using circular dichroic X-ray resonant magnetic scattering (CD-XRMS) as a function of the incident X-ray angle, we explore the depth evolution of chiral spin textures in two samples with different strengths of magnetocrystalline anisotropy. 
One FePd thin film with very high magnetocrystalline anisotropy, characterized by $Q_{PMA}=1.8$, exhibits an unexpectedly large Néel contribution. Angular-dependent CD-XRMS directly reveals a smooth transition from a hybrid Bloch-Néel chirality within the upper FePd layer towards a purely Néel-type structure at the lower FePd/Pd interface.
In the second FePd sample, despite a still relatively large $Q_{PMA}=1.45$, the domain walls were found to be purely Néel type. 
Combined with circular dichroic scanning transmission X-ray microscopy (CD-STXM), CD-ptychography, and macroscopic characterization of the structural order, magnetic properties, and surface morphology, we relate these observations to differences in the structural quality of the L1$_0$ phase of FePd.
These results indicate a crucial role of the long-range structural order in determining the formation of the magnetic structure.

\end{abstract}
\maketitle
\renewcommand{\thefigure}{\arabic{figure}}
\section{Introduction}
\par

Magnetic domain walls featuring chiral spin structures in ferromagnetic thin films exhibiting perpendicular magnetic anisotropy (PMA) have attracted considerable attention because of their unique interactions with spin currents and their relevance to modern spintronic technologies, including magnetic recording, logic, and memory devices~\cite{Foerster2016,Yang2020,Navas2014}. Devices based on the motion of magnetic domain walls can provide efficient low-power, high-velocity information transport \cite{Emori2013, Navas2014}. 
Recent studies on magnetic thin films hosting spiral magnetic textures discovered a broad range of emergent phenomena, including topological-chiral interactions~\cite{Grytsiuk2020}, coupling to materials with strong spin-orbit interaction~\cite{Skoropata2020}, and interactions with superconductors~\cite{Robinson2010}. These make ferromagnetic thin films with PMA and chiral spin structures highly promising as hybrid heterostructures for use in energy-efficient spintronic devices~\cite{Navas2014}.


PMA plays a dual role in such systems. While it stabilizes the out-of-plane magnetization required for perpendicular recording media, it can also suppress interface-driven magnetic textures such as closure domains, which themselves are promising information carriers for next-generation memory and neuromorphic computing concepts~\cite{Jiang2017,Yang2020}. 
Furthermore, high PMA provides stable, narrow domain walls~\cite{Blundell}, reduces depinning current densities~\cite{Navas2014}, and enhances current-driven domain-wall mobilities~\cite{Davydenko2025,Emori2013}. 
Thus, understanding the interconnection between PMA and the domain wall chirality is crucial for an efficient use in spintronic applications. In the present work, we explore the evolution from Bloch- to Néel-type domain walls and their chirality as a function of depth in FePd thin films with different strengths of PMA.

\begin{figure*}[t]
\includegraphics[width=\textwidth]{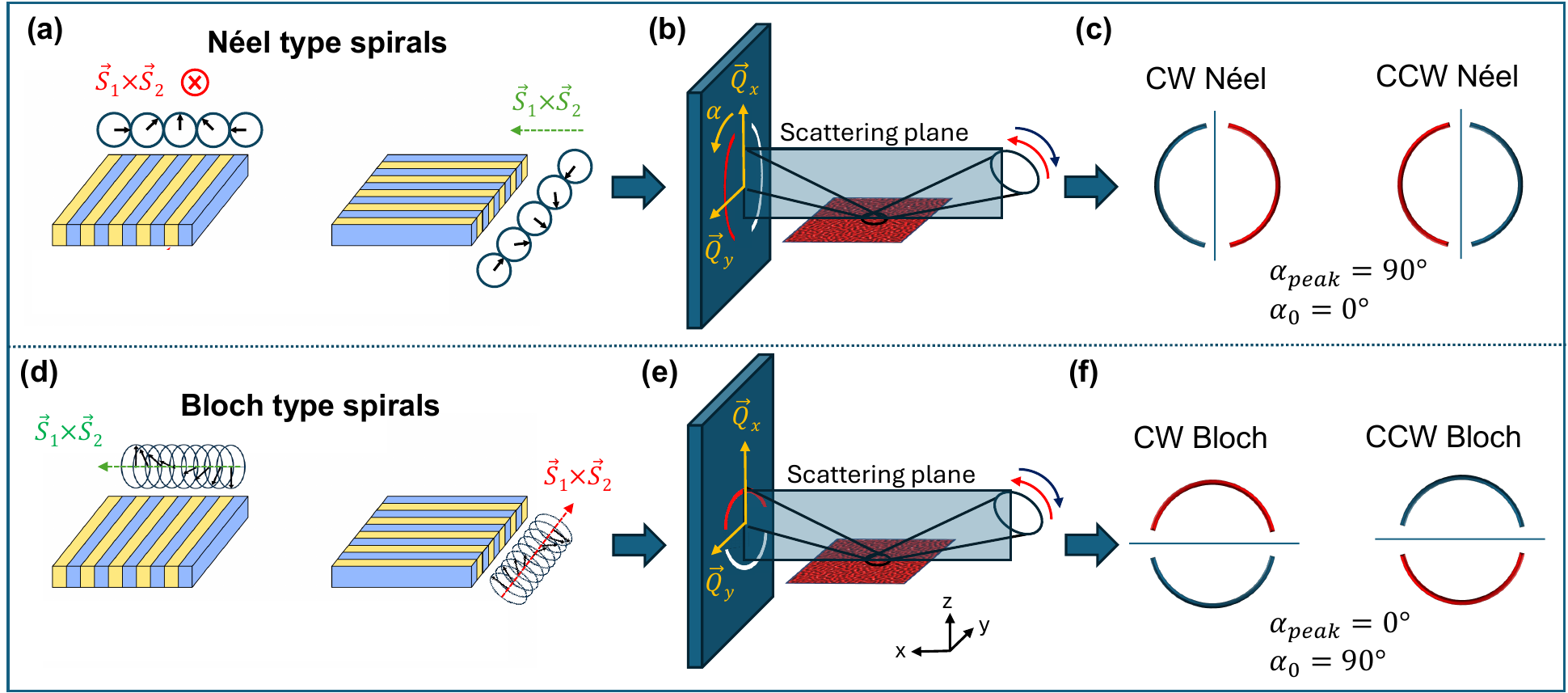} 
\caption{Schematics of the geometrical considerations for (a-c) Néel-type and (d-f) Bloch type chiral structures~\cite{Chauleau2018}: Néel-type chiralities obtain a rotational sense as schematically outlined in (a). Together with the measurement geometry (b), this leads to intensity minima (blue) and maxima (red) at $\alpha_{peak}=90^{\circ}$ and zero-crossing points at $\alpha_{0}=0^{\circ}$ on the detector as outlined in (c), with opposite intensity distributions for clock-wise versus counter-clockwise Néel chiralities. Bloch-type chiralities (d) in this measurement geometry (e) lead to minima and maxima at $\alpha_{peak}=0^{\circ}$ and zero-crossing points at $\alpha_{0}=90^{\circ}$ on the detector (f). Chiral structures that can be detected in this geometry are marked green, while chiralities with $\vec{c} \parallel \vec{y}$ cannot be detected in this setup and are marked red, see (a,d).}
\label{fig:CD-XRMS-geometric}
\end{figure*}

In perpendicularly magnetized systems, cycloidal Néel-type and helicoidal Bloch-type domain walls represent the two fundamental configurations, and are schematically displayed in Fig.~\ref{fig:CD-XRMS-geometric} (a,d). Their formation is governed by the interplay between perpendicular magnetic anisotropy, dipolar interactions, interfacial effects, and the Dzyaloshinskii-Moriya interaction (DMI)~\cite{Franke2021}.
The magnetic structure of domain walls in thin films with PMA is often considerably more complex than a simple Bloch- or Néel-type configuration. Kittel~\cite{Kittel1946} predicted that domain walls in thin films with PMA may consist of Bloch walls in the film interior with Néel caps forming near the surfaces in order to minimize stray-field energy. This has been experimentally confirmed in low-PMA FePd~\cite{Duerr1999,Navas2014}, as well as in CoCrPt thin films with PMA ~\cite{Navas2014}. D\"urr \textit{et al.}~\cite{Duerr1999} demonstrated the existence of Néel-type closure domains near the surfaces of low-PMA FePd thin films using circular dichroism X-ray resonant magnetic scattering (CD-XRMS). Navas \textit{et al.}~\cite{Navas2014} showed by polarized neutron reflectometry that, depending on the film thickness and PMA, such thin films host Bloch-like walls in the center of the film together with Néel caps at the surfaces and intermediate transition regions. 

Deviations from a simple picture of Bloch domain walls in the film interior and Néel-type closure domains at the interfaces have been predicted by Legrand \textit{et al.}~\cite{Legrand2018}, with Néel-to-Bloch contributions depending on the DMI. However, anisotropic DMI is strongly influenced by both the magnetocrystalline anisotropy and structural defects~\cite{Franke2021, Carvalho2023, Stellhorn2026}. In such systems, a direct observation of depth-dependent chiral magnetic domain walls is challenging and yet missing. \\


In this paper, we employ CD-XRMS to investigate the magnetic domain wall structure as function of depth inside two FePd thin films with $L1_0$ structural ordering and different strength of PMA. 
In particular, we will use the results of Chauleau \textit{et al.}~\cite{Chauleau2018}, who demonstrated that the shape of the circular dichroism pattern in CD-XRMS in reflection geometry can be used to unambiguously identify both the type and chirality of domain walls, schematically shown in Fig.~\ref{fig:CD-XRMS-geometric} .
L1$_0$-ordered FePd thin films exhibit out-of-plane magnetic domains characterized by alternating magnetization perpendicular to the sample surface.
This arises from strong uniaxial PMA energy ($E_{\mathrm{MA}}$), a consequence of the symmetry inherent to the $L1_0$ crystal structure~\cite{Laughlin2005}.
The PMA gives rise to an easy magnetization axis along the $<$001$>$ direction, with Bloch-type (helical) domain walls within the film interior, depending on the magnitude of the strength of the magnetocrystalline anisotropy \cite{Blundell, Navas2014}. For magnetic flux closure, this leads to Néel-type (cycloidal) closure domains near the film surfaces. Combining CD-XRMS at constant incident angle, and micromagnetic modelling \cite{Laan2003, Dudzik2000}, it has been found that the ratio between shape anisotropy and the PMA governs the ratio between Bloch- vs. Néel-type domain walls in FePd: low PMA leads to large Néel-type closure domains~\cite{Laan2003}, while they have been predicted to not be stable under strong PMA. \\


Micromagnetic simulations of magnetic textures that are strongly coupled to crystal structure require precise knowledge of the underlying crystallinity.
While Chauleau and Legrand \textit{et al.}~\cite{Chauleau2018, Legrand2018} have investigated the CD-XRMS pattern at a fixed incident angle and evaluated the depth structure using micromagnetic simulations, here we probe the depth-dependent Bloch and Néel-type domain wall formation experimentally by using CD-XRMS as a function of the incident X-ray angle. We combine our findings with observations from CD-based Scanning Tunneling X-Ray Micropscopy (CD-STXM) and CD-Ptychography. We correlate the magnetic domain wall structure with the observed structural properties, influenced by grain boundaries and terrace-like surface features.\\

$\,$\\

\section{Methods}\label{kap:Theory-domains}
\subsection{Sample growth and characterization}
\subsubsection{Growth of FePd thin films}
\par  L1$_0$-ordered FePd thin films with space group symmetry P4/mmm were grown following the protocol described in Refs.~\cite{DoktorarbeitGehanno, Gehanno1997, Gehanno1999}. Two samples, \textbf{FePd-high} and \textbf{FePd-mid} with different magnetocrystalline anisotropies, controlled by different FePd growth modes and thicknesses, were prepared using a state-of-the-art molecular beam epitaxy system (DCA Instruments, Finland) operated at ultra-high vacuum ($10^{-10}$ mbar):
\begin{itemize}
    \item \textbf{FePd-high}: 
    Pd (1.5\,nm) / FePd (38\,nm) / Pd (47\,nm) / Cr (4\,nm) / MgO(001)\\
    \item \textbf{FePd-mid}: 
     Pd (2\,nm) / FePd (70\,nm) / Pd (70\,nm) / Cr (1\,nm) / MgO(001).
\end{itemize}
 For both samples a Cr seed layer was first deposited onto commercial MgO(001) substrates (MaTecK GmbH) to inhibit interdiffusion, followed by a thick Pd buffer layer to accommodate lattice mismatch. To promote high structural ordering for the growth of the FePd layer, the Pd/Cr/MgO stack was annealed at 500\,K prior to the co-deposition of Fe and Pd. 
 
\par \textbf{FePd-mid} was grown with two subsequent growth modes to achieve a slightly lower magnetocrystalline anisotropy of the FePd layer, while keeping a high structural long range order of the L1$_0$-ordered phase consisting of alternating Fe and Pd monolayers. First there was a shuttered growth at room temperature, alternately opening the Fe and Pd shutters, followed by codeposition of Fe and Pd at 500~K~\cite{Stellhorn2019}, both with an FePd thickness of 35~nm. In total, this results in the higher FePd layer thickness of 70~nm. A thin Pd capping layer was deposited on top of the FePd to prevent surface oxidation for both \textbf{FePd-high} and \textbf{FePd-mid} thin films.
\\

\subsubsection{X-ray diffraction}
\par The structural quality of the thin films was verified by X-ray diffraction (XRD) measurements using a Bruker AXS D8 Advance system. Figure~\ref{fig:chiralities}(e) shows XRD measurements of the FePd (001) and (002) Bragg peaks (with $2\theta$(\textbf{FePd-high})=$49.8^{\circ}$ and $2\theta$(\textbf{FePd-mid})=$49.25^{\circ}$ of the (002) peak positions). These (002) Bragg peak positions correspond to out-of-plane lattice constants of $c_{\perp}$(\textbf{FePd-high})$=$3.66$\mathrm{\AA}$, and $c_{\perp}$(\textbf{FePd-mid})$=$3.7$\mathrm{\AA}$. Considering that L1$_0$-ordered FePd in the space group P4/mmm has a smaller out-of-plane lattice constant compared to 
disordered FePd (fcc-phase) \cite{DoktorarbeitGehanno, icsd-Ichitsubo}  (see Table \ref{tab:Ku-table}), this indicates different structural qualities and long-range orders of the L1$_0$-phase in the two samples.
The measured values are listed together with comparisons from \cite{DoktorarbeitGehanno, icsd-Ichitsubo} in Table \ref{tab:Ku-table}.

The long-range structural order $S$ of the L1$_0$-phase varies between $S$=$0$ for a nonstoichiometric phase to $S$=$1$ for a completely ordered phase \cite{Warren1969, DoktorarbeitGehanno} and impacts on the integrated intensities of the Bragg reflections, $A_{001}$ and $A_{002}$, via the structure factors $FF^*_{001}$ and $FF^*_{002}$:
\begin{eqnarray}
\frac{A_{001}}{A_{002}} = \frac{LP(\theta_{001})FF^*_{001}(S)\sin(\theta_{002})}{LP(\theta_{002})FF^*_{002}\sin(\theta_{001})}.
\label{eq:M_s}
\end{eqnarray}
The dependence of $FF^*_{002}(S)$ on $S$ is explained in detail in Refs.~\cite{Warren1969, DoktorarbeitGehanno, PhDStellhorn}, $LP(\theta)$ describes the Lorentz-polarization factor of the instrument at an incident angle $\theta$.
A perfect L1$_0$-phase yields $S=1$ with finite (001) Bragg peak appearance, while lower structural order is connected to $S<1$ and lower (001) Bragg intensities, e.g.~due to mixed fcc and L1$_0$-phases, atomic intermixing, or structural defects, and depends strongly on the growth conditions. Here, the samples show a structural order of $S(\textbf{FePd-high})=0.63\pm0.02$ and $S(\textbf{FePd-mid})=0.31\pm0.04$. In comparison, Gehanno \emph{et al.}~\cite{Gehanno1997} reported a value of $S=0.8$ for high quality L1$_0$-phase leading to strong PMA. 

\subsubsection{Magnetization measurements}

\begin{figure}
	\includegraphics[width=1\linewidth]{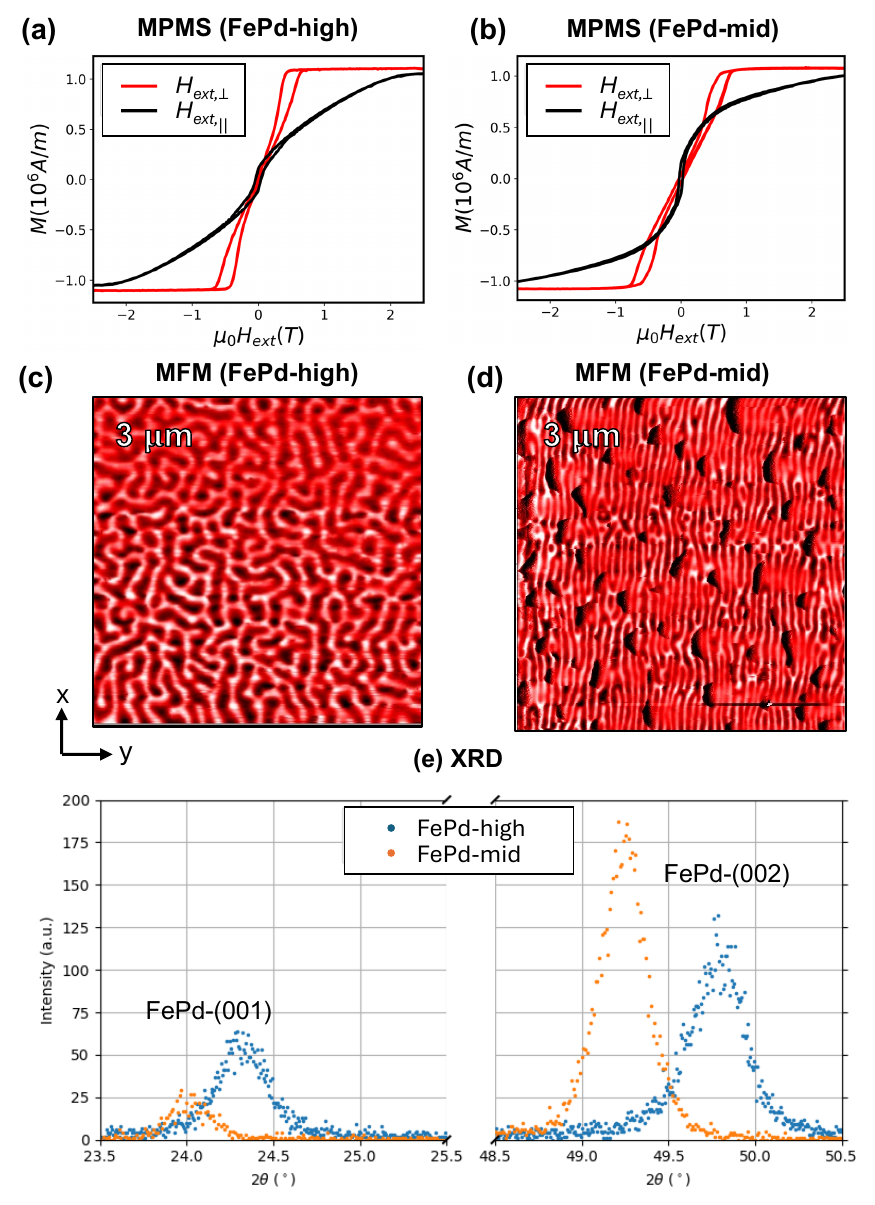}
	\caption{ Magnetic and structural characteristics of samples \textbf{FePd-high} and \textbf{FePd-mid}: (a,b) Magnetization measurements using an MPMS system (c,d) Lateral magnetic domain pattern as measured by MFM, and (e) XRD measurements of the FePd(001) and FePd(002) Bragg peaks. All measurements were performed at room temperature.}
	\label{fig:chiralities}
\end{figure}

\begin{table*}[htbp]
    \centering

    \label{tab:Ku-table}
    \begin{tabular}{@{}l c c c c c@{}}
        \toprule
        & $K_u$ ($\text{Merg/cm}^3$) & $Q_{\text{PMA}}$ & $D$ (nm) & $S$ & $c_{\perp}$ (\AA) \\
        \midrule
        \multicolumn{6}{@{}l}{\textit{Measurements}} \\
        \addlinespace[2pt]
        \textbf{FePd-high} & $14.43 \pm 0.05$ & $1.8 \pm 0.05$  & $70 \pm 2$ & $0.63 \pm 0.02$ & 3.66 \\
        \textbf{FePd-mid}  & $12.77 \pm 0.05$ & $1.45 \pm 0.03$ & $58 \pm 2$ & $0.31 \pm 0.04$ & 3.7  \\
        \midrule
        \multicolumn{6}{@{}l}{\textit{Literature}} \\
        \addlinespace[2pt]
        L1$_0$-FePd \cite{DoktorarbeitGehanno}  & 10.3 & 1.55 & $\approx 60$ & 0.79 & 3.714 \cite{icsd-Ichitsubo} \\
        disordered FePd \cite{DoktorarbeitGehanno} & 0.7  & 0.11 & ---          & 0    & 3.807 \cite{icsd-Ichitsubo} \\
        \bottomrule
    \end{tabular}
    \caption{Magnetic and structural parameters of \textbf{FePd-high} and \textbf{FePd-mid}: Magnetocrystalline anisotropy $K_u$, Quality factor $Q_{\text{PMA}}$, Domain width $D$,  Structural order $S$, and out-of-plane lattice constant $c_{\perp}$}

\end{table*}

\par Average bulk magnetization as a function of applied magnetic field, $M(H)$, was measured using a magnetic properties measurement system (rf SQUID-MPMS) from Quantum Design at the Juelich Research Center (JCNS-2) in Germany.  
Fig.~\ref{fig:chiralities}(a,b) shows the magnetization curves measured at room temperature with magnetic fields applied parallel ($H_{ext,||}$) and perpendicular ($H_{ext,\perp}$) to the thin film surface.  
From the ratio of the integrated $M(H_{ext,\perp})$ and $M(H_{ext,||})$ hysteresis loops, the magnetocrystalline anisotropy constant, $K_u$, can be extracted as follows \cite{Gehanno1997}:
\begin{eqnarray}
\int_0^{M_s}{(H_{ext,\perp}-H_{ext,||})dM} = K_u + K_{sh},
\label{eq:M_s}
\end{eqnarray}
where $M_s$ is the saturation magnetization and the shape anisotropy is $K_{sh}=-\frac{1}{2}\mu_0M_s^2$.
The ratio $K_u / K_{\mathrm{sh}}$  
yields the quality factor $Q_{PMA}$, and describes the relative strength of PMA compared to shape anisotropy. High PMA is characterized by values $Q_{PMA}>1$, whereas $Q_{PMA}<1$ indicates the higher impact of the shape anisotropy with an in-plane easy magnetization axis, which is described for L1$_0$-ordered FePd thin films in detail in~Ref.~\cite{Stellhorn2019}.  The strength of PMA of both samples has been evaluated to be $Q_{PMA}(\textbf{FePd-high})=1.8\pm0.05$ and $Q_{PMA}(\textbf{FePd-mid})=1.45\pm0.03$.
\subsection{Microscopy measurements}	
\par Structural and magnetic surface properties have been investigated at zero magnetic field and at room temperature by atomic and magnetic force microscopy (AFM, MFM) using an Agilent 5400 microscope in magnetic ac mode. Figure~\ref{fig:chiralities}(c,d) shows MFM of \textbf{FePd-high} and \textbf{FePd-mid} reflecting their conditions during the CD-XRMS measurements: a maze domain structure with interlaced domains in \textbf{FePd-high} (Fig. \ref{fig:chiralities}(c)), while \textbf{FePd-mid} exhibits an aligned magnetic domain pattern (Fig. \ref{fig:chiralities}(d)). The domain widths extracted by measuring the average peak-to-peak domain distance of the MFM images are $D(\textbf{FePd-high})=70\pm2$~nm and $D(\textbf{FePd-mid})=58\pm2$~nm. 

The PMA, magnetic domain width and lateral patterning, as well as the structural order $S$, are linked via crystal-field interactions and the spin-orbit coupling \cite{Coey2010}:
the strength of PMA impacts on the magnetic domain width \cite{Gehanno1997, DoktorarbeitGehanno}, correlating larger magnetic domain widths with stronger PMA. The lateral domain pattern is maze-like for high PMA, whereas a parallel aligned stripe pattern can only be achieved in samples of lower PMA \cite{Laan2003}. In short, the structural quality of FePd impacts on the PMA. 
The higher PMA, higher domain width, and higher structural order of \textbf{FePd-high} is confirmed by its maze domain pattern, whereas \textbf{FePd-mid} with a slightly lower PMA has a smaller domain width, lower structural order, and a laterally parallel stripe pattern.

\subsection{Circular-Dichroism scanning transmission X-ray microscopy and Circular-Dichroism Ptychography}\label{supplptycho}

\par Both circular-dichroism scanning transmission X ray microscopy (CD-STXM) and CD-ptychography image the magnetic domain structure (available from the circular-dichroic component) after transmission of the X-ray beam through the thin film samples at normal incidence. We have used an X-ray energy at the Fe L$_3$ resonance condition ($E=707\,$eV) and have calculated the asymmetry ratio as described in the Appendix 
to retrieve the purely magnetic domain pattern.

\subsection{Circular-Dichroism X-ray Resonant Magnetic Scattering}

\par Circular-Dichroism X-ray Resonant Magnetic Scattering (CD-XRMS) measurements of \textbf{FePd-high} and \textbf{FePd-mid} were conducted at two different beamlines: I10 with the RASOR endstation RASOR  at the Diamond Light Source, UK, and  BL29 using the X-ray scattering endstation MaReS \cite{BL-29} at the ALBA Synchrotron, Spain. Both experiments were performed at room temperature and zero magnetic field at the Fe L$_3$ resonance condition ($E = 707\,\mathrm{eV}$), where the magnetic signal is strongly dominated by the iron contribution. Off-resonance scans at $E = 690\,\mathrm{eV}$ (at Diamond) and $E = 700\,\mathrm{eV}$(at ALBA) were also performed for comparison. The scattering geometry is shown in Fig.~\ref{fig:CD-XRMS-geometric}(b).\\

\par At I10, \textbf{FePd-high} was investigated. The sample-to-detector distance was set to $sdd_{I10}$=138\,mm, and the X-ray beam had a size of $200\times200$\,\textmu m$^2$. With a sample size of \textbf{FePd-high} with $7\times10\,\mathrm{mm}^2$, no footprint correction for sample illumination was necessary. To define the magnetic field history, the sample was first saturated out-of-plane in a magnetic field exceeding $1\,\mathrm{T}$, followed by a ramp-down to zero field.
Incident angles in the range $\theta_{in} = 10^\circ$–$70^\circ$ were used which includes maximum attenuation lengths greater than 60\,nm~\cite{Henke-tables}. A transformation of $\theta_{\text{in}}$ into the X-ray attenuation length and a comparison with the probed depth of the FePd thin films was performed using the Henke tables~\cite{Henke-tables}. In \textbf{FePd-high}, the lower FePd interface to the Pd buffer layer is reached at $\theta_{\text{in}}=40^{\circ}$.

\par At BL-29, \textbf{FePd-mid} was investigated. The scattering endstation at BL29 has a sample position with a sample-to-detector distance $sdd_{BL29}$=402\,mm and slit sizes between $5\times5$\textmu m$^2$ - $50\times50$ \textmu m$^2$. Here, incident angles of $\theta_{in} = 10^\circ$–$40^\circ$ have been investigated. Unfortunately, it was not possible to probe the lower FePd/Pd interface in \textbf{FePd-mid} due to the 70 nm thickness of this layer.

\par At both beamlines and at each angle, measurements were carried out using both left- and right-circularly polarized light. By selecting specific polarization channels before and after scattering, different magnetization components could be probed. Magnetic moments lying in the scattering plane, $\vec{M}_{\parallel}$, are detected in the $\sigma \rightarrow \pi'$ and $\pi \rightarrow \sigma'$ channels, while magnetization components perpendicular to the scattering plane, $\vec{M}_{\perp}$, are accessed via the $\pi \rightarrow \pi'$ channel~\cite{McMorrow1996, Laan2003}. Using circularly polarized light, which couples to the complex magnetic vector $\vec{M}_{\pm} = \vec{M}_{\parallel} \pm i \vec{M}_{\perp}$, allows us to probe sinusoidal, helical, or otherwise chiral magnetic structures such as Bloch and Néel walls ~\cite{Laan2003, Hannon1988}. \\

For details on the recalculation of the detector image to $I(|Q_x, Q_y|)$, error function treatment, and the asymmetry ratio calculation, see Appendix.

\subsection{Identification of Bloch- and Néel-type domain wall chirality from CD-XRMS}
Chauleau and Legrand \textit{et al.}~\cite{Chauleau2018, Legrand2018} demonstrated that a direct experimental observation of the shape of the circular dichroism scattering pattern from CD-XRMS in reflection geometry can be used to determine both the type and chirality of domain walls. Magnetic domains with a scattering vector $|\vec{Q}|$ generate a circular scattering pattern, centered around the specular reflection. This $|\vec{Q}|$
defines the average magnetic periodicity, which represents the full spatial repetition period of alternating domains and domain walls, here $2D = 2\pi / \vert{}\vec{Q}\vert{}$.
Figure~\ref{fig:CD-XRMS-geometric} displays the typical dichroic scattering pattern for Néel (a-c) and Bloch (d-f) domain walls: a ring-like scattering pattern with $|\vec{Q}|$ corresponding to the periodicity of the domains and domain walls, and an intensity distribution following the extinction rules for a dichroic scattering pattern from chiral magnetic structures along the ring, i.e.~following the azimuthal angle $\alpha$ \cite{ShileiZhang2018}. 
Néel-structures show a scattering pattern with opposite intensity minima and maxima (sketched red and blue, respectively) on the horizontal, at $\alpha_{peak}=90^{\circ}$, and intensity zero-crossings at $\alpha_{0}=0^{\circ}$. In contrast, helical Bloch walls propagating along $\vec{y}$ produce the opposite pattern, with intensity peaks along the vertical, at $\alpha_{peak}=0^{\circ}$ and zero-crossings along the horizontal at $\alpha_{0}=90^{\circ}$. Clockwise (CW) and counterclockwise (CCW) chiral structures produce inverted scattering patterns, mirrored at the vertical (Néel) or horizontal (Bloch) lines. 
A hybrid chiral structure exhibiting a depth-dependent ($\theta_{in}$-dependent) change between Bloch and Néel spirals is characterized by a change of $\alpha_0$ and the position of peak intensity on $I(\alpha)$.
Even for a maze-like domain structure exhibiting a distribution of in-plane oriented domains, the symmetry of the CD-XRMS scattering pattern provides a clear fingerprint that distinguishes between Néel- and Bloch-type spirals oriented along the sample surface~\cite{Chauleau2018}.\\

\section{Results and Discussion}\label{kap:results}

\begin{figure*}
\includegraphics[width=1\linewidth]{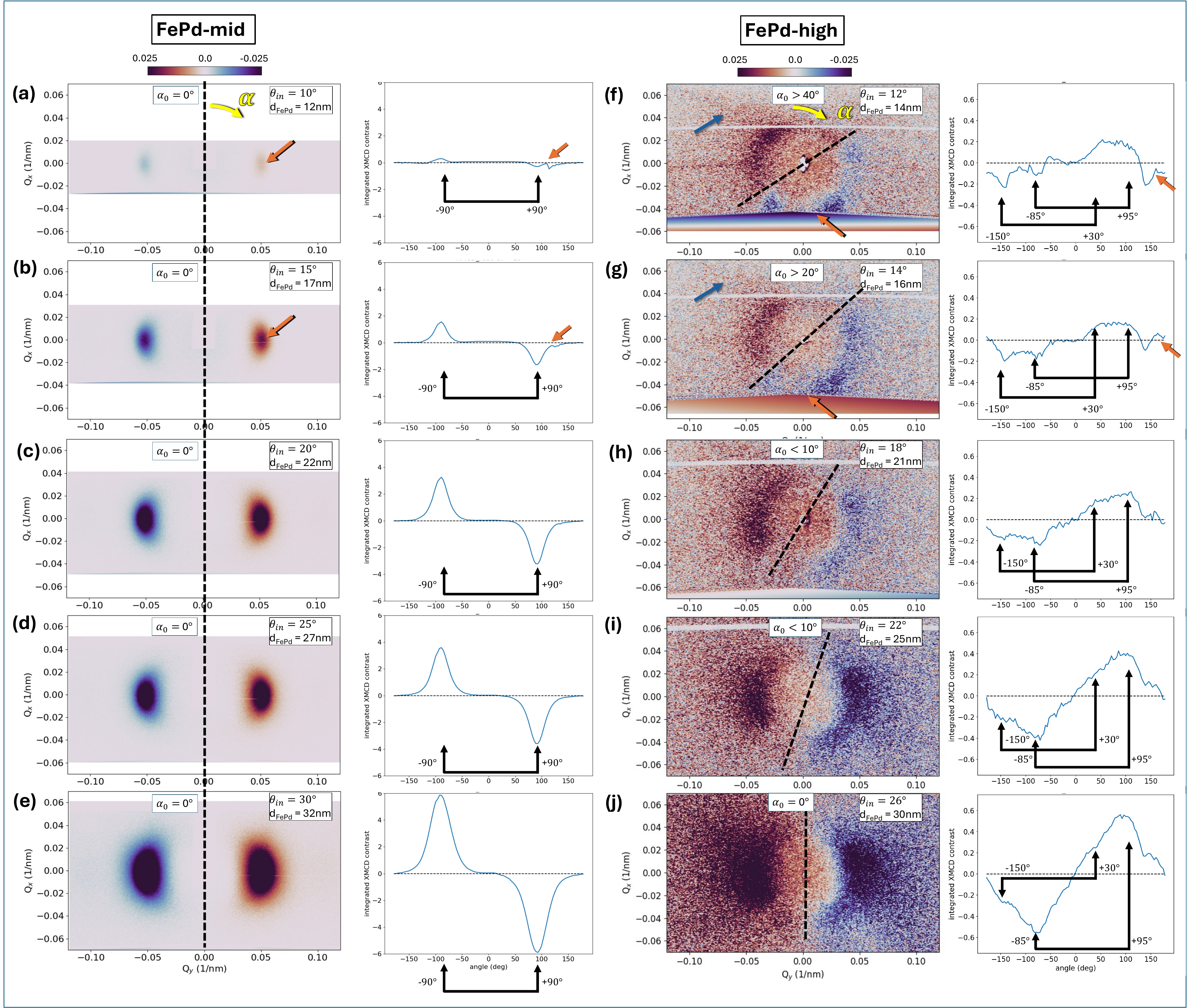}
\caption{2D CD-XRMS patterns (left) and extracted $I(\alpha)$ (right) of (a-e) \textbf{FePd-mid} and (f-j) \textbf{FePd-high} for selected $\theta_{\text{in}}$. The yellow arrow indicates the direction of $\alpha$, with $\alpha_0$ marked by a dashed black line. Orange arrows mark the impact of artifacts on the lineplots of $I(|Q_x, Q_y|, \alpha)$ (see Appendix
). Blue arrows mark high-intensity regions at $\alpha=45^{\circ}$ connected to the surface structure of FePd as outlined in the discussion.}
\label{fig:CD-XRMS-detector}
\end{figure*}

In order to investigate the type and depth-dependence of chiral magnetic domain walls in L1$_0$-ordered FePd thin films, we performed CD-XRMS measurements as a function of incident angle. Figure~\ref{fig:CD-XRMS-detector} shows the CD-XRMS measurements collected on 
\textbf{FePd-mid} (a-e) and \textbf{FePd-high} (f-j) at selected incident angles between $\theta_{in}=10-30^{\circ}$, covering a depth-range from 12-32$\,$nm. The black dashed line indicates the azimuthal orientation of $\alpha_0$, the intensity zero-crossing. The radius $|\vec{Q}|$ of the ring-like scattering cross-section $I(|Q|, \alpha)$ defines the domain and domain-wall periodicity, where $|Q|$ is the wavevector transfer, and $\alpha$ is the azimuthal angle around the specular peak.
From the scattering ring, the average $|\vec{Q}|$ is determined as $|\vec{Q}|=(0.042-0.05\,\mathrm{nm}^{-1})\pm0.003\,\mathrm{nm}^{-1}$, varying slightly as function of depth inside FePd. This variation in the average $|\vec{Q}|$ arises from the broadening of the Bragg peaks in $I(|\vec{Q},\alpha|)|$ at higher $\theta_{in}$ (see Appendix 
Fig.~\ref{fig:App_Broad_peaks}). This $|\vec{Q}|$ arises from the magnetic domains and domain walls in FePd and matches the domain periodicity of 120-140~nm observed by MFM in Fig.~\ref{fig:chiralities}(c,d). The data reduction process from raw images to the extracted 2D $I(Q_x, Q_y)$ and the according 1D $I(\alpha)$ at constant $|Q|$, as well as measurement artifacts and their impact, are described in the Appendix
. From Fig.~\ref{fig:CD-XRMS-detector} we evaluate the magnetic chiral nature and depth-dependence of both samples.\\

For \textbf{FePd-mid}, the 2D CD-XRMS patterns in Figs. \ref{fig:CD-XRMS-detector}(a-e) show a constant vertical mirror symmetry (see black dashed line) and constant $\alpha_{0}=0^{\circ}$ for each incident angle, with sharp intensity maxima (red signal in Fig. \ref{fig:CD-XRMS-detector}(a-e)), and minima  (blue signal in Fig. \ref{fig:CD-XRMS-detector}(a-e)), at $\alpha_{peak}=\pm90^{\circ}$, respectively, 
and increasing intensity as function of depth. 
To accurately identify $\alpha_{peak}$ and $\alpha_0$, we integrate the intensity on the scattering ring as function of $|Q|$. 
The resulting 1D $I(\alpha)$ unambiguously identify $\alpha_{peak}=\pm90^{\circ}$ and $\alpha_{0}=\pm0^{\circ}$ for each incident angle, which indicate in-plane Néel-type chiral domain walls with constant chiral direction over the whole probed depth of \textbf{FePd-mid}.\\


For \textbf{FePd-high}, the 2D CD-XRMS patterns show a rotation of the mirror symmetry and $\alpha_{0}$ as function of incident angle in Figs.~\ref{fig:CD-XRMS-detector}(f-j), marked by the black dashed lines. Accordingly, $\alpha_{peak}$ rotates as function of the incident angle, which is evaluated in the 1D line-plots: two $\pi$-shifted peak combinations outlined by black arrows are observed  (i) at ($-85^{\circ},+95^{\circ}$), which increases in intensity as function of depth, and (ii) at ($-150^{\circ},+30^{\circ}$), which stays constant in intensity as function of depth. For (i), a slight shift from the typical Néel-type response at $\alpha=\pm90$ indicates truncated Néel-type spirals, e.g. from partial magnetic components out of the propagation plane of the Néel spiral. These features are named "Truncated-Néel" (Tr-Néel) structures. The signature (ii) at ($-150^{\circ},+30^{\circ}$), indicates the additional existence of a hybrid Bloch-Néel-spiral, as proposed by~\cite{Legrand2018}, with values of $\alpha_{0}>40^{\circ}$. As a function of incident angle, $\alpha_{0}$ rotates monotonically to $\alpha_{0}=0^{\circ}$, i.e., a Néel-type vertical symmetry.
Considering that horizontal mirror symmetries denote a Bloch-type chirality, and vertical mirror symmetries denote a Néel-type chirality, this suggests a crossover from a hybrid Bloch-Néel-type spiral to a pure Néel-type spiral, as a function of depth inside the FePd layer.\\

\begin{figure*}
\includegraphics[width=\textwidth]{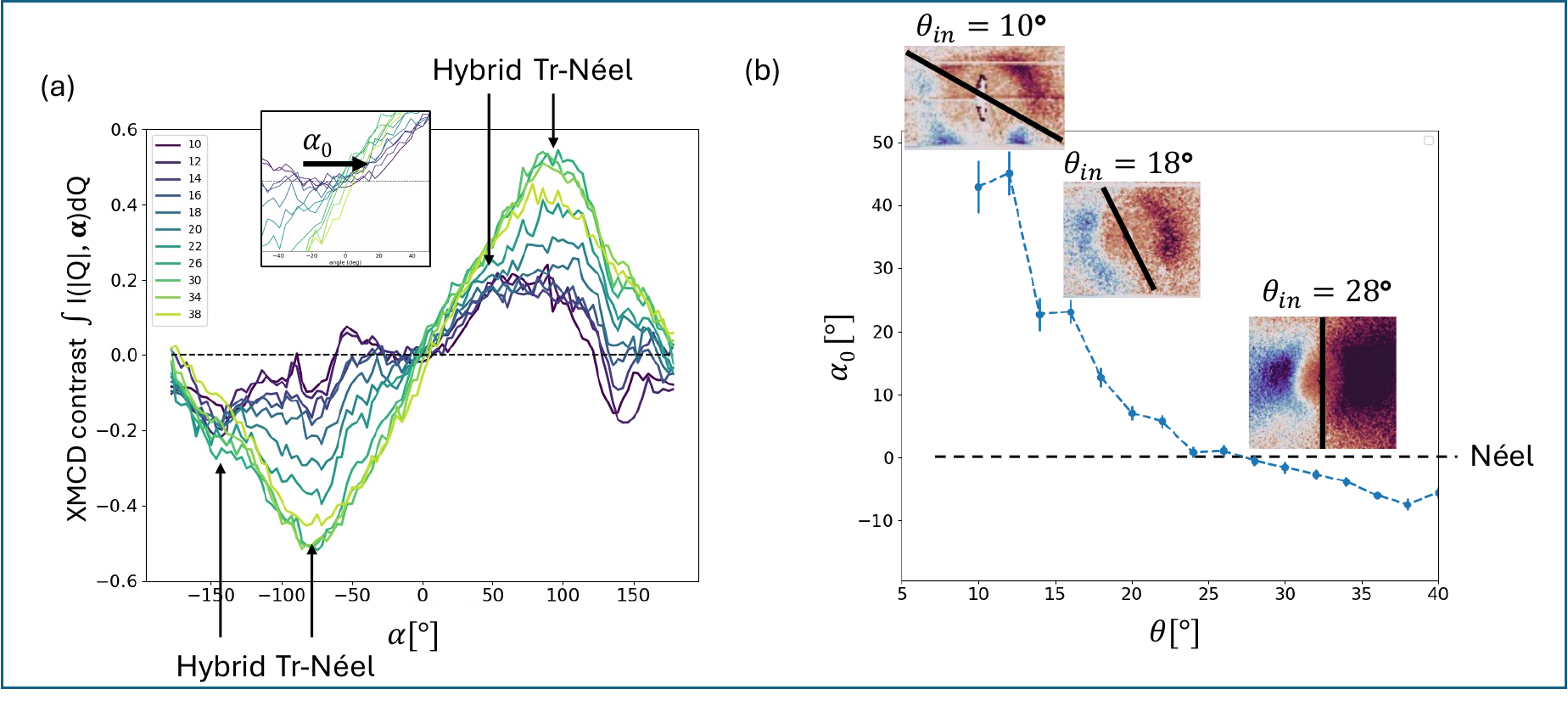}
\caption{(a) $I(|Q_x, Q_y|, \alpha)$ of \textbf{FePd-high} for selected $\theta_{\text{in}}$, arrows mark the positions of Néel- and Bloch-type contributions, and the inset indicates a shift of $\alpha_0$ with $\theta_{\text{in}}$. (b) Sinusoidal fit results of $\alpha_0$ to (a) as function of $\theta_{\text{in}}$. Additional insets visualize the rotation of $\alpha_0$ in the 2D scattering images with increasing $\theta_{\text{in}}$.}
\label{fig:zero-crossing}
\end{figure*}

Figure~\ref{fig:zero-crossing}(a) presents the lineplots $I(|Q|, \alpha)$ of \textbf{FePd-high} in Fig.~\ref{fig:CD-XRMS-detector}(a) for selected incident angles to superimpose the trend as function of incident angle. 
Corresponding to a hybrid Bloch-Néel-spiral, an according shift of the zero-peak position $\alpha_0$ is expected. A sinusoidal fit to each $I(|Q|, \alpha)$ was used to estimate $\alpha_0$, with results of $\alpha_0$ shown in Fig.~\ref{fig:zero-crossing}(b). 
A systematic shift from $\alpha_0 = -40^\circ$ to $\alpha_0 = 0^\circ$ with increasing $\theta_{\text{in}}$ is observed, as highlighted in the inset in Fig.~\ref{fig:zero-crossing}(a). Comparison with the two-dimensional scattering patterns (see insets in Fig.~\ref{fig:zero-crossing}(b)) confirms a rotation of $\alpha_0$. Altogether, this supports the existence of a hybrid Bloch-Néel-type spiral at low $\theta_{\text{in}}$ near the upper FePd layer surface, transitioning to pure Néel walls towards greater depth, i.e., near the FePd/Pd-buffer interface. Surprisingly, both samples exhibiting high PMA with $K_u>12$ Merg.cm$^{-3}$ show strong contributions from Néel-type closure domains at the FePd layer surfaces.\\

In addition to the scattering ring at $|\vec{Q}|=(0.042-0.05\,\mathrm{nm}^{-1})\,\pm\,0.003\,\mathrm{nm}^{-1}$, 
pronounced scattering intensity of the CD-XRMS patterns in Fig.~\ref{fig:CD-XRMS-detector} around $\alpha = 45^{\circ}$ has been observed, outlined by blue arrows at low $\theta_{in} < 18^{\circ}$. 

\begin{figure*}
    \includegraphics[width=0.7\textwidth]{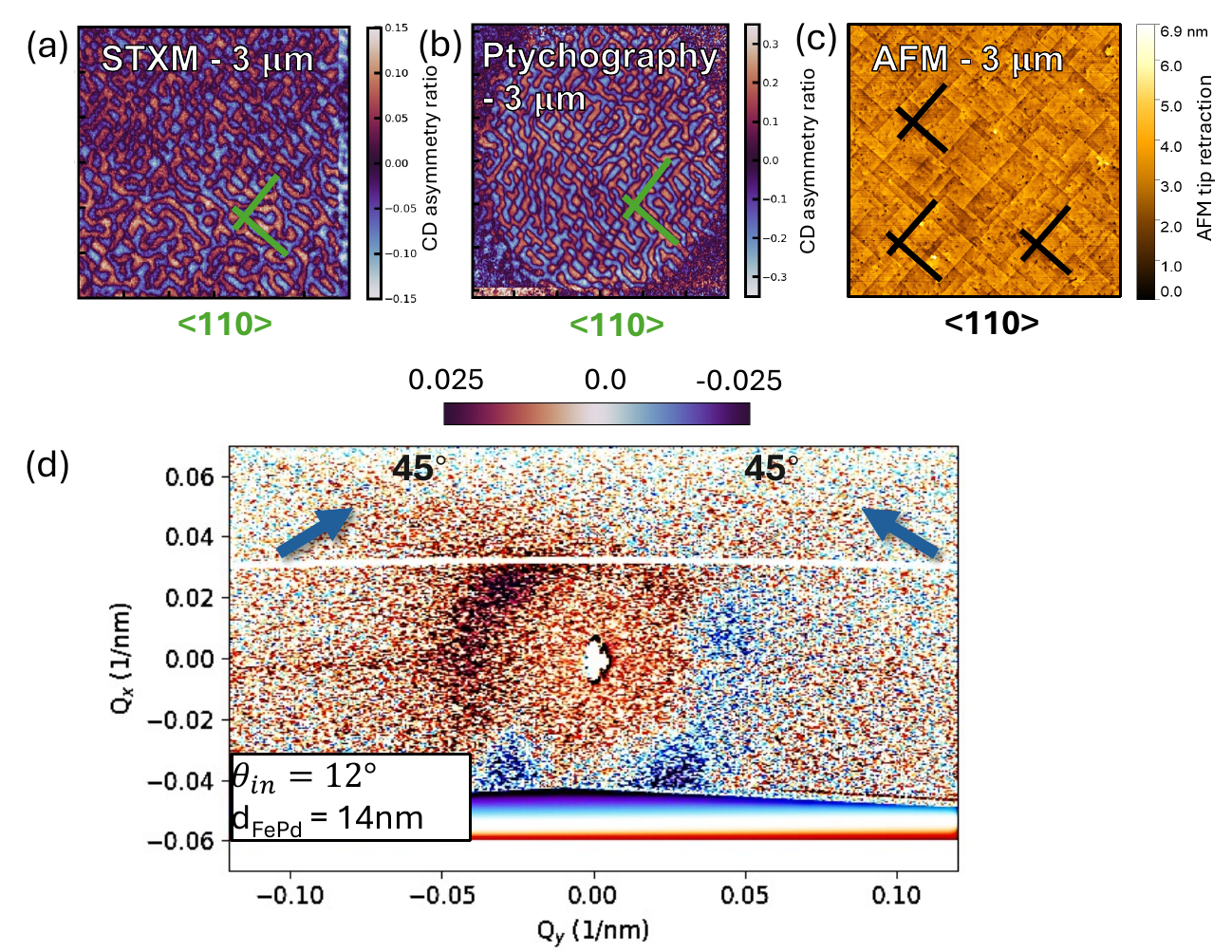}
    \caption{3x3 \textmu m images of a comparable high-PMA FePd sample, displaying (a) the CD-STXM asymmetry ratio, (b) the CD-Ptychography asymmetry ratio, and (c) the FePd layer surface structure as measured by AFM. (d) shows an exemplary CD-XRMS scattering cross section at $\theta_{in}=12^{\circ}$ with artifacts at $\alpha=45^{\circ}$ (marked by blue arrows), connected to the surface structure. Insets mark the direction of $<$110$>$ oriented surface terraces.}
    \label{fig:Ptycho}
\end{figure*}

This is seen in all measurements on \textbf{FePd-high}, in both resonant and off-resonant ($E_{\text{off}} = 690\,\text{eV}$) conditions, with strongest intensity at low $\theta_{in} < 18^{\circ}$, but persistent through all measured scattering angles. 
The persistence of this feature off resonance indicates a structural origin. 
To investigate this, CD-STXM and CD-Ptychography of a similar sample to \textbf{FePd-high} were performed to obtain an image of the magnetic domain structure. The sample used Was grown following the same method as \textbf{FePd-high}, but with a thicker FePd layer (57~nm), resulting in a slightly larger magnetocrystalline anisotropy constant, $K_u=16.5$ Merg.cm$^{-3}$.
The results are displayed in Fig.~\ref{fig:Ptycho}(a,b). They are compared with the FePd surface structure as measured by AFM in Fig.~\ref{fig:Ptycho}(c). STXM provides a pixel-by-pixel transmission image of the sample, and the CD-asymmetry ratio for magnetic domain imaging could still contain residual structural artifacts due to spatial drift or sharp sample boundaries (grain boundaries, edges, terraces). Ptychography uses a larger probe and collects diffraction patterns with overlapping positions on the sample. From the set of diffraction patterns, a high-resolution complex-valued real space image of the sample can be calculated using an iterative phase retrieval algorithm \cite{Thibault2012, Thibault2016}. This method decouples structural amplitude and phase contrast from magnetic dichroism and eliminates optical diffraction artifacts, yielding quantitative, high-resolution magnetic domain images free from structural edge artifacts.



A preferred lateral direction of magnetic domains along the $<$110$>$ crystalline directions can be observed, marked by green lines in Fig.~\ref{fig:Ptycho}(a,b)). A comparison with AFM in Fig.~\ref{fig:Ptycho}(c) shows the same directional ordering of a surface terrace structure, marked by orange lines. These terraces lead to scattering streaks within the $Q_y$-$Q_x$ maps with an inclination angle of $\alpha = 45^{\circ}$, for comparison shown in enlarged in Fig.~\ref{fig:Ptycho}(d), marked by blue arrows.

Typically, in high-PMA FePd with long-range structural order, grain boundaries can lead to the observed surface terraces and are a sign of high-quality epitaxial L1$_0$-ordered FePd \cite{Halley2002}, such as \textbf{FePd-high}. 

\newpage

\section{Conclusion}\label{sec:conclusion}

We compare the depth-dependent chiral magnetic domain-wall textures of two L1$_0$-FePd thin films and relate the spin textures observed by CD-XRMS to their structural and magnetic properties. Our observations show that these thin films, despite their large magnetocrystalline anisotropy, exhibit a surprisingly high proportion of Néel walls, although they are expected to display predominantly Bloch-type domain walls with only minor Néel-type closure-domain contributions at the sample surfaces. These observations are strongly correlated with the long-range structural order and magnetization of the films.\\

\textbf{FePd-high}, with $Q_{PMA}(\textbf{FePd-high})=1.8\pm0.05$, exhibits a high degree of structural L1$_0$ order, resulting in high PMA and large magnetic domain widths, with a maze-like magnetic domain pattern. In addition, it exhibits grain boundaries, surface terraces, and an according alignment of the magnetic domains along these structural features.
Such a FePd thin film is expected to exhibit strong Bloch-type contributions to the domain-wall structure \cite{Navas2014}.
However, CD-XRMS reveals that \textbf{FePd-high} exhibits a hybrid Bloch--Néel spiral with overlapping chiral components at the upper FePd surface, evolving into purely Néel-type walls toward the lower FePd/Pd interface.
Concerning the depth dependence of the magnetic structure, the crossover from a hybrid Bloch--Néel spiral to a purely Néel-type spiral occurs at $\theta_{in}=24^{\circ}$, corresponding to a probing depth of approximately 25\,nm.
Overall, the \textbf{FePd-high} sample has a total thickness of 38\,nm, consisting of approximately 25\,nm of hybrid Bloch--Néel spirals and 13\,nm of purely Néel-type spirals located near the lower FePd/Pd buffer interface, corresponding to 34\,\% of the total sample thickness.\\

\textbf{FePd-mid}, with $Q_{PMA}(\textbf{FePd-mid})=1.45\pm0.03$, exhibits slightly lower long-range structural order and PMA, which is expected to be associated with larger Néel closure domains compared with~\textbf{FePd-high}~\cite{Laan2003}.
This is reflected in a virgin-state magnetic stripe-domain pattern, in contrast to the maze-domain structure observed in the higher-PMA sample (see Fig. \ref{fig:chiralities}(c,d)). Using CD-XRMS, we show that \textbf{FePd-mid} supports purely Néel-type chiral domain walls throughout the entire FePd layer thickness, despite its relatively large $Q_{PMA}=1.45$. \\


For comparison, both \textbf{FePd-high} with $K_{u} = 14.43\pm0.05$~Merg.cm$^{-3}$, and \textbf{FePd-mid} with $K_{u} = 12.77\pm0.05$~Merg.cm$^{-3}$ exhibit substantially larger magnetocrystalline anisotropy constants than a comparable CoCrPt thin film with a maze-domain configuration ($K_{u} = 1.225$~Merg.cm$^{-3}$ for a 20~nm film) reported in Ref.~\cite{Navas2014}. In that study, an overall contribution of approximately 25\% from purely Néel-type closure caps at the film surfaces was observed. Given the significantly higher PMA of both FePd samples, a reduced Néel-type contribution would generally be expected. Therefore, the pronounced Néel-type character observed in \textbf{FePd-high}, as well as the purely Néel-type domain walls in \textbf{FePd-mid} without detectable Bloch-wall contributions, is unexpected. These findings indicate that future theoretical investigations of the DMI including structural dependencies such as grain boundaries and surface terraces, are essential for understanding the origin of the observed domain-wall configurations.\\


The origin of the preferred handedness of observed Bloch-type domain walls is ascribed to the existence of DMI with preferential direction driven by structural defects with a preferred orientation, interdiffusion, or grain boundaries, as discussed in~\cite{Stellhorn2026}. An in-plane preferential direction of the magnetic domains along the $<$110$>$ directions is observed by STXM and ptychography in Fig. \ref{fig:Ptycho}. This corresponds to the orientation of surface terraces as observed by AFM in Fig.~\ref{fig:Ptycho}(c), which originate from grain boundary growth along $<$111$>$ \cite{Halley2002}. Such a preferential orientation can additionally influence the preferred handedness of Néel and Bloch spirals with in-plane propagation. In addition, direct contact between FePd and the underlying Pd buffer layer can induce non-negligible spin-orbit interactions that distort spin spirals. \\

The presented results highlight the importance of detailed structural and depth-sensitive magnetic studies, as differences in the domain wall configurations can arise even in nominally similar systems. The observed behavior suggests that subtle changes in microstructure or interfacial effects, such as induced DMI, play a decisive role in stabilizing different chiral spin textures. Our findings challenge the conventional expectation that systems with strong PMA predominantly host Bloch-type spin spirals without preferred handedness and only surface-near contributions from Néel closure domains. This strongly impacts their application in spintronic devices, in which a tuning of the chiral magnetic states requires detailed knowledge about both the lateral and depth-dependent magnetic structure.
Future work will aim to directly correlate CD-XRMS measurements and domain wall chirality with DMI effects introduced by structural imperfections, complemented by first-principles calculations and micromagnetic simulations. The insertion of a non-magnetic interlayer between FePd and Pd would help to confirm or rule out artifacts caused by the Pd buffer layer.

\section{Acknowledgements}
The authors would like to thank Diamond Light Source for beamtime, and the staff of beamline I10 for measurements under two proposal rounds: 31510-1, and 35196-1. The experiments at the BOREAS beamline at ALBA Synchrotron were performed in collaboration with ALBA staff under the Proposal No 2022025733. 
We acknowledge MAX-IV for time at the Softimax beamline under Proposal No.~20230624.

A.S. acknowledges funding from the Tillväxtverket European Regional Development Fund project SREss3 (2020-2023). E.K. acknowledges the funding received from the European Union’s Horizon 2020 research and innovation programme under the Marie Skłodowska-Curie grant agreement No 101069104 — PRESSMAG — HORIZON-MSCA-2021-PF-01. D.V. and E.B. acknowledge support from the Swedish Research Council (Vetenskapsrådet) under Project Nos. 2018-04704 and 2021-06157. O.S. acknowledges financial support from the Swedish Foundation for Strategic Research (SSF) through SwedNess (grant no. GS1715-0008). M.V. acknowledges research grants PID2020‐116181RB‐C32 and PID2023-146354NB-C43 funded by MCIN/ AEI/10.13039/501100011033/.\\


\appendix
\renewcommand{\thesection}{} 
\makeatletter
\def\@seccntformat#1{}
\makeatother

\section{Data-processing}\label{kap:data-processing}
\renewcommand{\thefigure}{\arabic{figure}}

Data processing was performed as follows: for each incident angle, multiple snapshots were acquired and normalized by their respective counting times. The normalized snapshots were then averaged to generate a two-dimensional image per incident angle. To propagate uncertainties in the averaged dataset, the error for each pixel was calculated using
\[
\Delta I = \frac{\sum_N c_{\mathrm{norm}}(\theta)}{(N-1)N},
\]
where $c_{\mathrm{norm}}$ is the pixel-wise counting-time normalized count rate, and $N$ is the number of snapshots per incident angle $\theta$. 
To isolate the magnetic scattering contribution from charge scattering, the circular dichroism signal was computed using the asymmetry ratio in the main text named ``XMCD contrast''):
\[
I = \frac{I_l - I_r}{I_l + I_r},
\]
where $I_l$ and $I_r$ denote the intensities for left- and right-circularly polarized light after normalization and summing, respectively. To observe the full symmetry of the scattering pattern, measurements were performed without a beamstop, with reduced counting time per snapshot to avoid saturation of the CCD detector. For $\Delta I$ we have used the gaussian error propagation of pixel-wise error values originating from $I_l$ and $I_r$.

\begin{figure}[H]
\begin{center}
    \begin{minipage}{\textwidth}
        \centering
    \includegraphics[width=\textwidth]{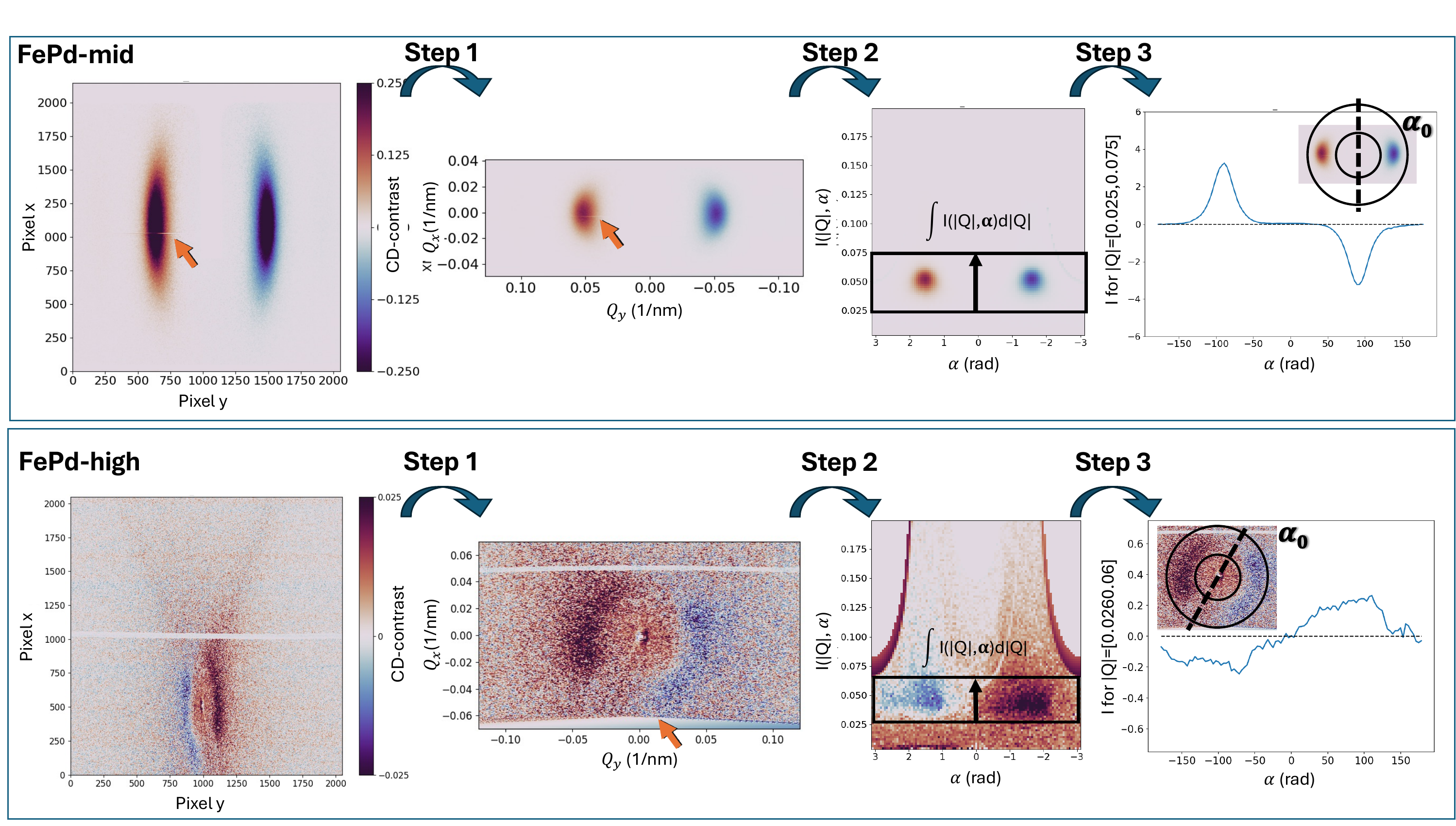}
    \caption{Sequential data analysis pipeline. Step 1: Ewald-corrected transformation from detector pixels to $Q$-space. Step 2: Recasting into $I(|Q|,\alpha)$ coordinates. Step 3: Radial $|Q|$-integration yielding 1D azimuthal $I(\alpha)$ distributions and sinusoidal extraction of the zero-angle rotation $\alpha_0$ (insets).}
    \label{fig:Data-analysis}
    \end{minipage}
\end{center}
\end{figure}

The subsequent data analysis steps for each incident angle are outlined in Fig. \ref{fig:Data-analysis}. Step 1: detector pixel coordinates were converted into reciprocal space vectors $Q$ by applying a projection correction to account for the Ewald sphere curvature on the flat detector, enabling extraction of the full scattering vector components $(Q_x, Q_y, Q_z)$. Step 2: Each image was then transformed into the representation $I(|Q|, \alpha)$, where $\alpha$ is the azimuthal angle around the specular peak. Step 3: This transformation was used to analyze the symmetry of the scattering pattern by integrating over a given range in $|Q|$, $\int(I(|Q|,\alpha)d|Q|)$, which yields $I(\alpha)$ as 1D lineplot. Step 3 is illustrated according to the reciprocal-space integration picture by the inset within the lineplots, showing the rotation of the zero-angle $\alpha_0$. $\alpha_0$ was extracted from the lineplots by fitting a sinusoidal function to the integrated data $I(\alpha)$.

\onecolumngrid
\twocolumngrid

\begin{figure}[H]
\begin{center}
\begin{minipage}{\textwidth}
        \centering
    \includegraphics[width=\textwidth]{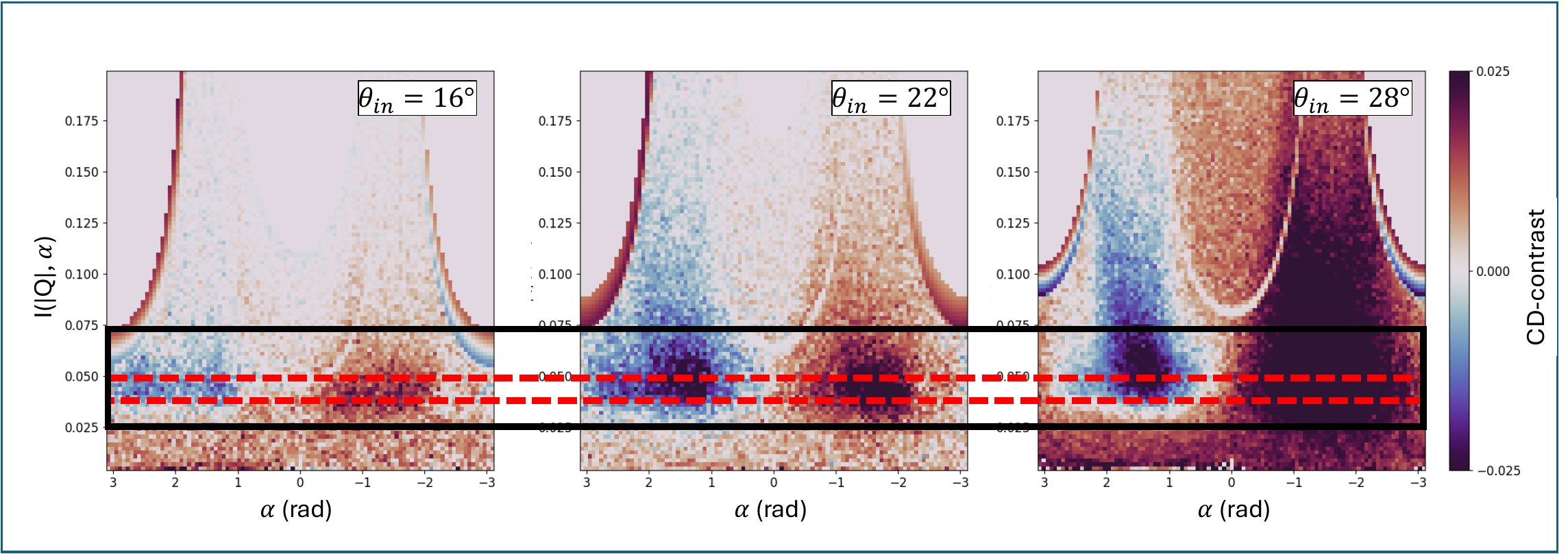}
    \caption{Integration range of $I(|\vec{Q}|,\alpha)|$ for obtaining 1D $I(\alpha)|$ lineplots (marked by the black line), and region of average $|\vec{Q}|=(0.042-0.05\,nm^{-1})$ (red dashed line).}
    \label{fig:App_Broad_peaks}
    \end{minipage}
\end{center}
\end{figure}

For \textbf{FePd-mid} and \textbf{FePd-high}, the integration was done over the ranges $d|Q|$(\textbf{FePd-mid}) = [0.025,0.075] and $d|Q|$(FePd-high) = [0.026,0.06], respectively, see Fig. \ref{fig:App_Broad_peaks} as example. For \textbf{FePd-high}, the scattering pattern at low incident angles interferes with the beam stop (white line in the images). This possibly leads to the striped zero-intensity-area of $I(Q_x, Q_y)$. Further artifacts are marked by orange arrows: In \textbf{FePd-high} the limited detector coverage (zero intensity at lower $Q_x$) impacts on $I(\alpha)$ at $\alpha>150^{\circ}$ (artifacts from these regions are largely avoided by limiting the upper $Q$-value in the integration), and in \textbf{FePd-mid} a sharp line of zero-intensity at $\alpha=90^{\circ}$. Measurements of \textbf{FePd-mid} were performed without a beamstop. It is important to note that the measured scattering intensity contains information from the sum of penetrated depths: at low $\theta_{in}$, mainly the thin film surface is probed, whereas at higher $\theta_{in}$, both the surface and bulk contribute to the observed scattering pattern. It should also be noted that minor variations of $|\vec{Q}|$ as a function of depth within the FePd layers are observed.\\

\onecolumngrid
\twocolumngrid

\bibliographystyle{apsrev4-2}
\bibliography{Bibo}

\end{document}